\documentclass[twocolumn,prl,aps,showpacs,amsmath,amssymb]{revtex4-1}
\usepackage{graphicx}
\usepackage{bm}
\begin{document}

\title{Suppression of viscous fluid fingering: a piecewise constant-injection process}
\author{Eduardo O. Dias}
\author{Fernando Parisio}
\author{Jos\'e A. Miranda}
\email[]{jme@df.ufpe.br}
\affiliation{Departamento de F\'{\i}sica, Universidade Federal de Pernambuco,
Recife, PE  50670-901 Brazil}


\begin{abstract}
The injection of a fluid into another of larger viscosity in a Hele-Shaw cell usually results 
in the formation of highly branched patterns. Despite the richness of these structures, in many practical 
situations such convoluted shapes are quite undesirable. In this letter we propose an efficient and 
easily reproducible way to restrain these instabilities based on a simple piecewise constant pumping protocol. 
It results in a reduction in the size of the viscous fingers by one order of magnitude.
\end{abstract}
\maketitle


Hydrodynamic fingering is one of a wider class of interfacial instabilities
occurring when one material is injected into another, or grows from a given chemical, biological or
geophysical process. Examples of such pattern forming systems include the classic Saffman-Taylor (ST) 
instability~\cite{Saf}, the dynamics of chemically reacting fronts~\cite{Wit}, the
growth of filamentary organisms~\cite{Go}, and lava flows~\cite{Grif}. These seemingly
unrelated processes produce complex patterns presenting similar morphological features, where
highly branched shapes and dendriticlike configurations may arise. In this context, the
control of the growth and form of the emerging complex morphologies has long been a
challenging topic, of great academic and technological relevance.

Due to its relative simplicity, and multiple applications the
ST instability has received much attention, and serves as a paradigm for pattern
formation systems~\cite{Rev}. This instability arises when a fluid is injected against a second of much larger
viscosity in the narrow gap between closely-spaced parallel glass plates of a Hele-Shaw (HS) cell~\cite{Hele}. Under
constant flow injection rate, the result is the development of vastly ramified interfacial
patterns~\cite{Pat,Gingra,Mir4,Praud,Li}. One of the most important practical situations related to this 
hydrodynamic instability is oil recovery~\cite{Gorell}, where petroleum is displaced by injection of water into 
the oil field in an attempt to extract more oil from the well. Interestingly, the dynamic behavior for flow in HS 
cells is described by the very same set of equations as those for flow in porous media~\cite{Rev}. In fact, the ST 
instability is a major source of poor oil recovery, once rapidly evolving fingers may reach the entrance of the well, and 
mainly water, and not oil is retrieved. This emblematic example clearly illustrates the importance of developing a fundamental 
understanding of the dynamics of this type of system, and to find ways to contain, and possibly suppress such interfacial 
disturbances.

Very recently, some research groups~\cite{Li2,Lesh,Mir10} have examined the possibility of avoiding the emergence of 
the usual branched morphology, by properly controlling the flow injection rate. Instead of considering a constant injection 
flux, they assumed a particular time-varying pumping rate which scaled with time like $C(n) t^{-1/3}$, where $C(n)$ depends 
on the interfacial wave number $n$. Their theoretical and experimental findings demonstrate that by using this specific pumping 
rate the formation of branched patterns is inhibited, and replaced by $n$-fold symmetric shapes. This process conveniently 
determines the number of emerging fingers. In spite of this controlling strategy, the resulting interfacial morphologies are notably 
noncircular, and still contain sizable fingers. Therefore, the interfacial fingering perturbations are not exactly wiped out, but 
redesigned into self-similar shapes with a prescribed number of fingers. In this sense, an efficient protocol for suppressing the 
development of the viscous fingering instability in radial HS flows is still lacking.

Differently to what is done in Refs.~\cite{Li2,Lesh,Mir10}, we consider a simple piecewise constant injection process which 
results in the actual suppression of the viscous fingering instability. In the usual constant
injection situation, a certain amount of fluid is pumped in a finite time, and interfacial fingering results. In the procedure we suggest 
the average injection rate is kept unchanged, so that the same amount of fluid is injected in the same time interval, but interfacial 
irregularities are restrained. 

We begin by briefly describing the traditional radial flow setup in confined geometry. Consider a HS cell of gap spacing $b$ 
containing two immiscible, incompressible, viscous fluids. The viscosities of the fluids are denoted as $\eta_{1}$ 
and $\eta_{2}$, and between them there exists a surface tension $\sigma$. Fluid 1 is injected into fluid 2 at a constant injection rate $Q_0$,
equal to the area covered per unit time. Linear stability analysis of the problem~\cite{Pat,Gingra,Mir4} considers harmonic distortions of 
a nearly circular fluid-fluid interface whose radius evolves according to ${\cal R}(\theta,t)= R(t) + \zeta_{n}(t) \cos{n \theta}$, 
where the time dependent unperturbed radius is $R(t)=R_{t}=\sqrt{R_{0}^{2} + Q_0t / \pi}$, $\theta$ represents the azimuthal angle, 
and $n$=0,$\pm 1$, $\pm 2$, $...$ are discrete wave numbers. The unperturbed radius of the interface at $t=0$ is represented 
by $R_{0}$, and the Fourier perturbation amplitudes are given by
\begin{equation}
\label{relax}
\zeta_{n}(t)=\zeta_{n}(0) \exp \{I_{0}(n)\}, ~~~~ I_{0}(n)= \int_{0}^{t} \lambda(n) {\rm d}t'
\end{equation}
where the linear growth rate is
\begin{equation}
\label{growth}
\lambda(n)= \left [ f(n)\frac{Q_0}{R^{2}} - g(n)\frac{1}{R^3}\right ]\;,
\end{equation}
%
%
%
with $f(n)=(A |n| - 1)/ 2 \pi$, $g(n)=[b^{2}\sigma|n| (n^{2} - 1)]/ [12 (\eta_{1} + \eta_{2})]$, and 
$A=(\eta_{2} - \eta_{1})/(\eta_{2} + \eta_{1})$ being the viscosity contrast. If $I_{0}(n) > 0$ the disturbance grows, indicating instability. 
In Eq.~(\ref{growth}) we notice opposing effects of the viscosity difference between the fluids (destabilizing) and of the surface tension 
(stabilizing). A relevant information can be extracted at the linear stage: the existence of a series of critical radii $R_{c}(n)$ [or critical 
times $t_c(n)$] at which the interface becomes unstable for a given mode $n$ [defined by setting $\lambda(n)=0$], characterizing a cascade of modes~\cite{Mir4}. 
Therefore, to write the linear solution in a more realistic way, one should consider the interval $[t_c(n),t_f]$ only, because integration over $[0,t_c(n)]$ would 
lead to an artificial diminishing in the size of the fingers, so that $\zeta_n(t_c(n))<\zeta_n(0)$. This is an unphysical effect because we assume that the amplitudes 
can not go below $\zeta_n(0)$ due to noncontrollable factors, e.g., irregularities on the surface of the plates~\cite{Gingra}. For this reason we will consider the initial perturbation 
to be independent of $n$, that is, $\zeta_n(0) \equiv \zeta_0$. Notice, however, that for lower modes $t_c(n) \ll t_f$, and solution (\ref{relax}) can be used as a good approximation.

We proceed by describing the stabilization protocol. Consider a radial HS flow, and suppose that one is required to pump a fluid into the cell, at a specified 
average rate $Q_0$, during the time interval $[0,t_f]$. This defines the final area occupied by the injected fluid, ${\cal A}_f=\pi R_0^2+Q_0 t_f$. Given the quantities ${\cal A}_f$ and 
$t_f$, our goal is to design a simple injection process that suppresses viscous fingering events occurring at usual constant pumping procedure which utilizes the same input parameters
${\cal A}_f$ and $t_f$. The results presented here refer to ${\cal A}_f \approx 140$ ${\rm cm}^2$ and $t_f=28.0$ s ($ Q_0=5.00$ cm$^2$/s) and to the following set of
characteristic physical parameters: $R_0=0.30$ cm, $b=0.10$ cm, $\zeta_0=R_0/2400 \propto 10^{-4}$ cm, $\sigma=63.0$ dyne/cm, $\eta_2=5.21$ g/cm s, and $\eta_1 \approx 0$. These values are consistent 
with those used in typical experimental realizations~\cite{Pat,Gingra,Mir4,Praud,Li}. We chose $A \approx 1$ because in this limit the most {\it unstable} 
situation is reached, so that the results can only improve for $0 < A < 1$.

\begin{figure}
\includegraphics[width=2.0 in]{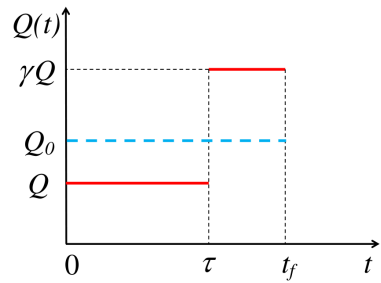}
\caption{(Color online) Injection rate $Q(t)$ as a function of time for a two-stage piecewise constant protocol. The equivalent constant injection rate $Q_{0}$ is represented 
by the horizontal dashed line.}
\label{injection}
\end{figure}

Out of a plethora of possibilities for time-varying injection scenarios, the simplest nontrivial procedure is a two-stage piecewise constant pumping (Fig.~\ref{injection}), whose choice 
will be justified shortly. Specifically, we split the time interval $[0,t_f]$ into $[0,\tau]$, during which the constant injection rate is $Q$; and $[\tau, t_f]$, 
with $t_f=(1+\beta)\tau$, for which the injection is given by $\gamma Q$. The requirement that the average injection remains unchanged demands that
\begin{equation}
\label{injection_def}
Q=\frac{(1+\beta)}{(1+\gamma \beta)}Q_0\;,
\end{equation} 
with $\beta>0$ and $\gamma \beta > -1$. If the parameter $\gamma$ is negative we have an injection stage followed by a period of suction. For $\gamma>0$ two nontrivial 
possibilities arise: if $\gamma<1$ the injection in the first stage is stronger than in the second, and for $\gamma>1$ the weaker injection stage comes first. For $\gamma=1$ 
we recover the usual constant pumping situation.

As commented earlier, the formal output of a linear analysis is given by Eq. (\ref{relax}) for the constant injection, while for the lower modes in the two-stage pumping we 
have $\zeta'_n(t)=\zeta_0\exp \{I'(n)\}$ with
\begin{eqnarray}
\nonumber
I'(n)= \int_{0}^{\tau} \lambda_1(n) {\rm d}t+\int_{\tau}^{t_f} \lambda_2(n) {\rm d}t=
2 \pi f(n) \Lambda (R_f,R_0) \\
-\frac{2 \pi g(n)(1+\beta \gamma)}{Q_0 (1+\beta)}\left[ \frac{1}{\gamma} \Gamma (R_f,R_{\tau})+\Gamma (R_{\tau},R_0)\right]\;,
\label{I1}
\end{eqnarray}
where $\lambda_1$ and $\lambda_2$ refer to the first and second stages with their respective injection rates, 
$\Lambda(x,y)= \ln (x/y)$, and $\Gamma(x,y)=(1/y - 1/x)$. Note that while $R_0$ and $R_f$, the initial and final 
unperturbed radii, are constant, the radius at $t=\tau$, is a function of the free parameters, $R_{\tau}=R_{\tau}(\beta,\gamma)$.

If we are to suppress instabilities, at least for the initial modes, we must impose that 
${\zeta'}_n/{\zeta}_n=\exp\{I'-I_{0}\}<1$ in the end of the whole process, i.e., at $t=t_f$.  
By analyzing the overall sign in the argument of the exponential one shows that the only 
scenario that leads to stabilization is that with $\gamma>1$, or, an initial stage with a 
relatively weak injection rate followed by a stronger (by a factor of $\gamma$) injection stage. 
This can be understood by recalling that the lower the mode the larger its maximum growth rate.
So, to suppress the initial and larger instabilities a slow injection is needed. 
For the other scenarios no values of $\beta$ and $\gamma$ produce a stabilizing effect.
We have also checked oscillating pumping with $Q(t)>0$ and oscillations involving injection and 
suction in each cycle, both resulting in an enhancement of the ST instability.
Thus, our task is to establish a general and simple way to obtain 
optimal values, $\beta^*$ and $\gamma^*$, within the selected scenario ($\beta>0$ and $\gamma>1$).

It is a key point to realize that during the piecewise constant injection the cascade of modes 
occurs differently. In general, the ``wakening" of lower modes still occurs in the first stage of pumping. However, 
for $t=t_f/(1+\beta)=\tau$ the injection rate jumps discontinuously  
and two effects occur: the most unstable modes prior to the sudden change in the pumping regime have their growth rates drastically 
decreased, and, at the same time, a certain number of new modes become unstable instantaneously. After that the cascade process 
continues. We have used a smoothed out version of the step-like injection function depicted in Fig.~\ref{injection} and verified that 
the outcome is virtually unchanged. This signifies that our results are not due to any peculiarity related to the discontinuous 
nature of $Q(t)$.

In our quest for the optimal parameters $\beta^*$ and $\gamma^*$, we try to devise simple rules that are {\it independent} of 
the discrete wave number. For this reason, we will suppose that the relevant modes approximately fit in one out of two classes: 
(i) modes that become unstable soon after the beginning of the first stage (lower modes), and (ii) modes which attain their regimes 
of instability exactly at $t=\tau$ or a bit later (higher modes). The remaining modes, i.e., those which become unstable at the end 
of first stage $[t \sim t_f/(1+\beta)]$ and in the final part of the second stage $(t \sim t_f$) are not critical because their 
regime of strong instability lasts for a short time. The effectiveness of these considerations will be made evident in what follows.
\begin{figure}[ht]
\includegraphics[width=2.3 in]{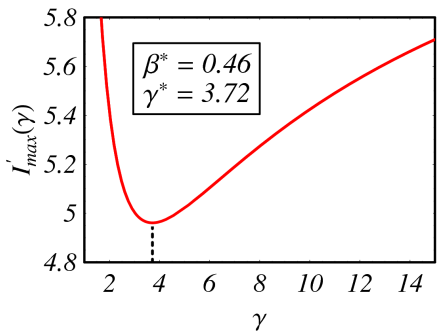}
\caption{(Color online) Behavior of $I'_{max}=I''_{max}$ as a function of $\gamma$. The absolute minimum at $\gamma^{*}=3.72$ is indicated.} 
\label{gamma}
\end{figure}
For the relevant modes of type (i) we have the perturbations appropriately described by Eq. (\ref{I1}), 
while for the modes of type (ii) we have $\zeta''_m(t)=\zeta_0\exp \{I''(m)\}$, where
\begin{eqnarray}
\nonumber
I''(m)= \int_{\tau}^{t_f} \lambda_2(m) {\rm d}t=
2 \pi f(m) \Lambda(R_f,R_0)\\
-\frac{2 \pi g(m)(1+\beta \gamma)}{\gamma Q_0 (1+\beta)}\Gamma(R_f,R_{\tau})\;.
\label{I2}
\end{eqnarray}
In order to get rid of the dependence on $n$ and $m$, we will focus on the modes associated to the largest perturbation 
amplitudes in each stage, $n_{max}$ and $m_{max}$, respectively. Therefore, we can guarantee that the size of the 
fingers corresponding to these wave numbers constitute an upper bound for the scale of all other modes.
This is obtained by setting $\partial I'/\partial n=0$ and $\partial I''/\partial m=0$, yielding
\begin{equation}
n_{max}=\sqrt{\frac{1}{3}+\frac{2 Q_0 (\eta_1+\eta_2) \gamma (1+\beta) \Lambda(R_f,R_0)}{ \pi b^2 \sigma (1+\beta \gamma)
\left[ \Gamma (R_f,R_{\tau})+\gamma \Gamma (R_{\tau},R_0)\right]}}\;,
\end{equation}
and
\begin{equation}
m_{max}=\sqrt{\frac{1}{3}+\frac{2 Q_0 (\eta_1+\eta_2) \gamma (1+\beta) \Lambda(R_f,R_{\tau})}{ \pi b^2 \sigma 
(1+\beta \gamma) \Gamma (R_f,R_{\tau})}}\;.
\end{equation}
\begin{figure}[hb]
\includegraphics[width=2.3 in]{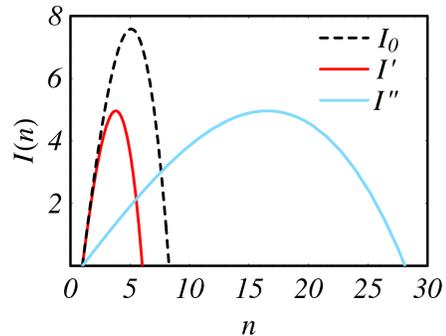}
\caption{(Color online) The continuous curves represent $I'$ and $I''$ as functions of $n$, calculated for 
$\beta^*$ and $\gamma^*$. The dashed curve shows the corresponding quantity $I_{0}$ for the constant injection process.} 
\label{bands}
\end{figure}
The referred upper bounds are then $I'(n_{max})$ and $I''(m_{max})$.
Before imposing any minimization condition on these quantities we make sure
that the suppression of the instabilities is uniform, with higher modes as controlled as the lower ones, 
by setting the constraint
\begin{figure}
\includegraphics[width=1.3 in]{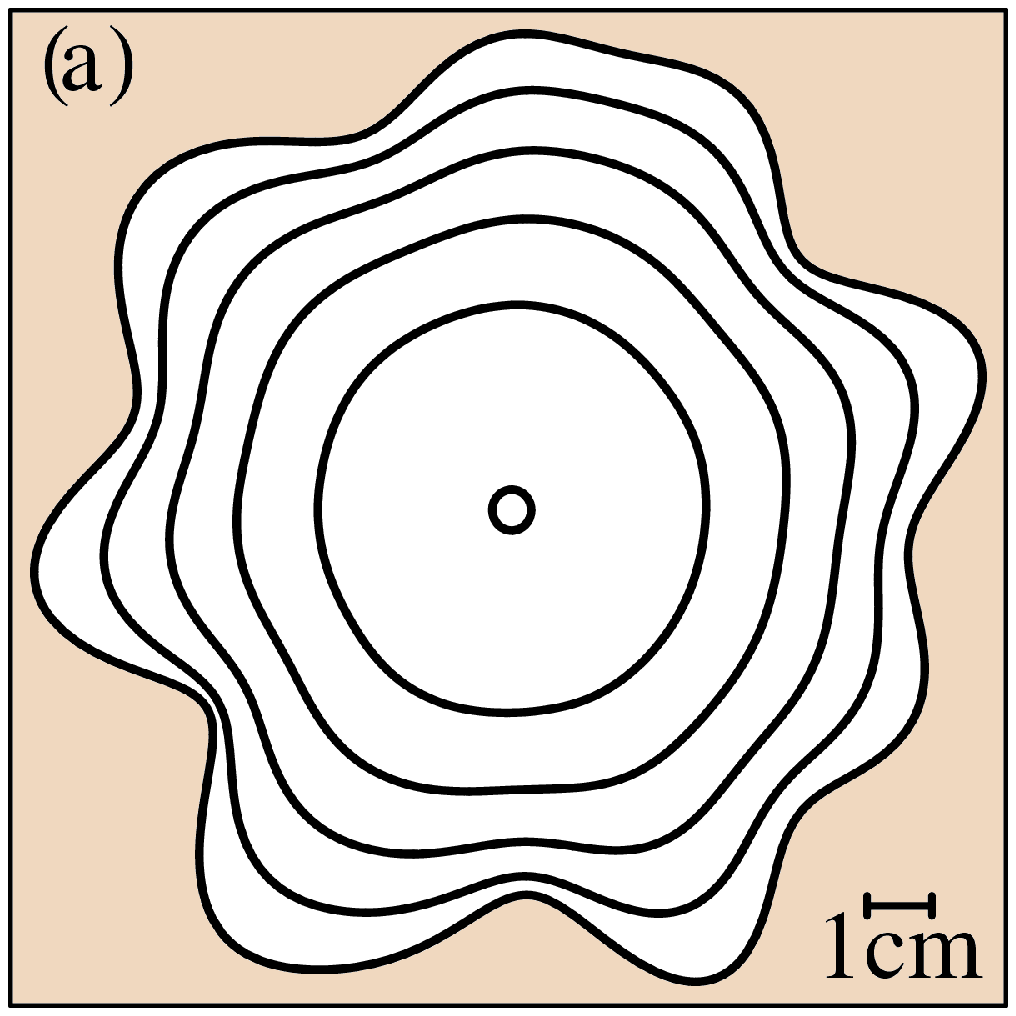}
\includegraphics[width=1.3 in]{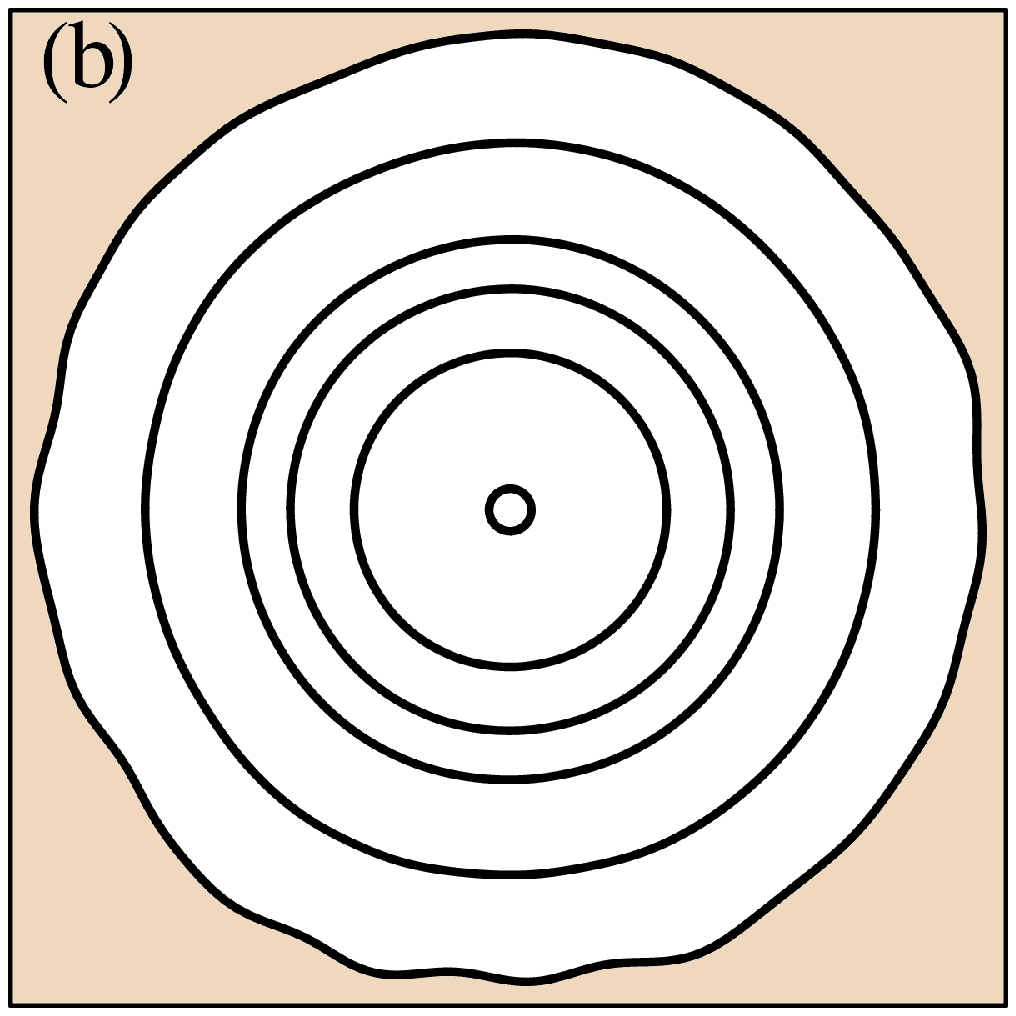}\\
\includegraphics[width=1.3 in]{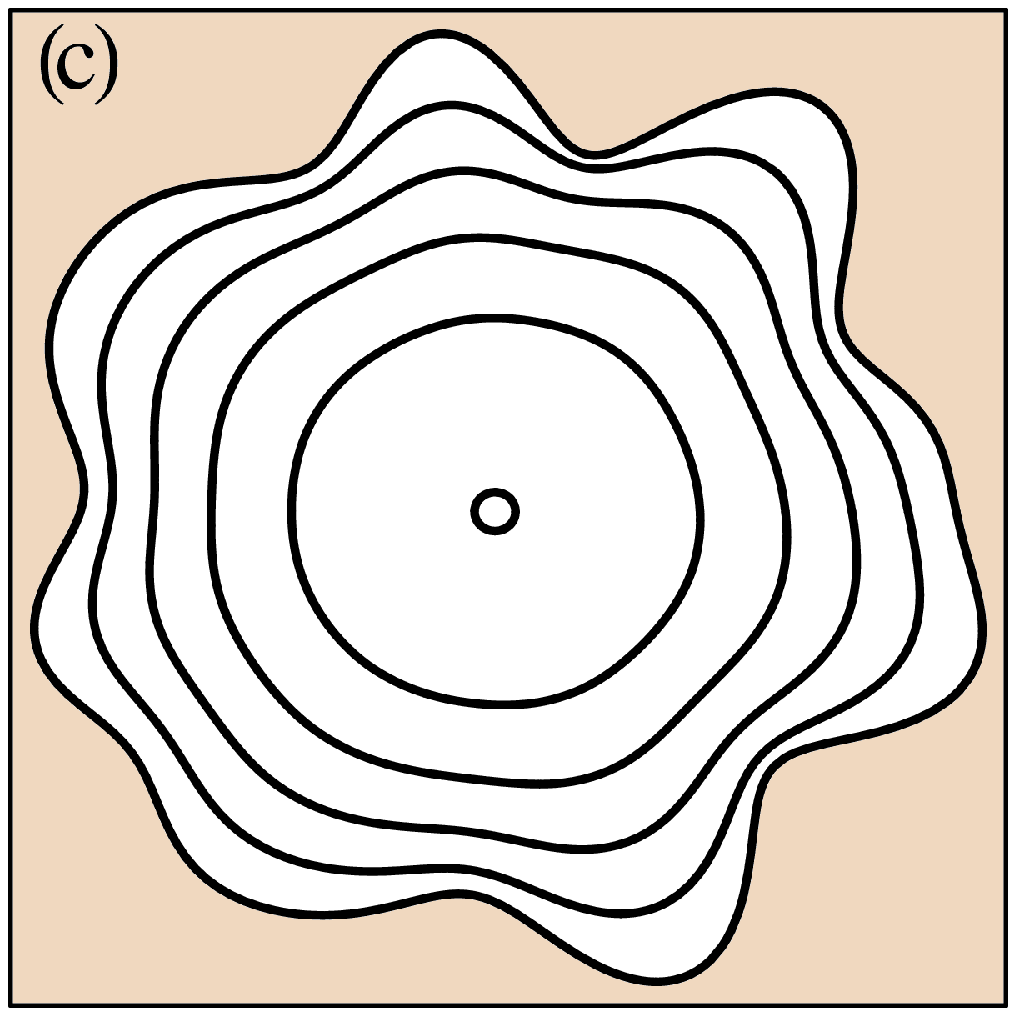}
\includegraphics[width=1.3 in]{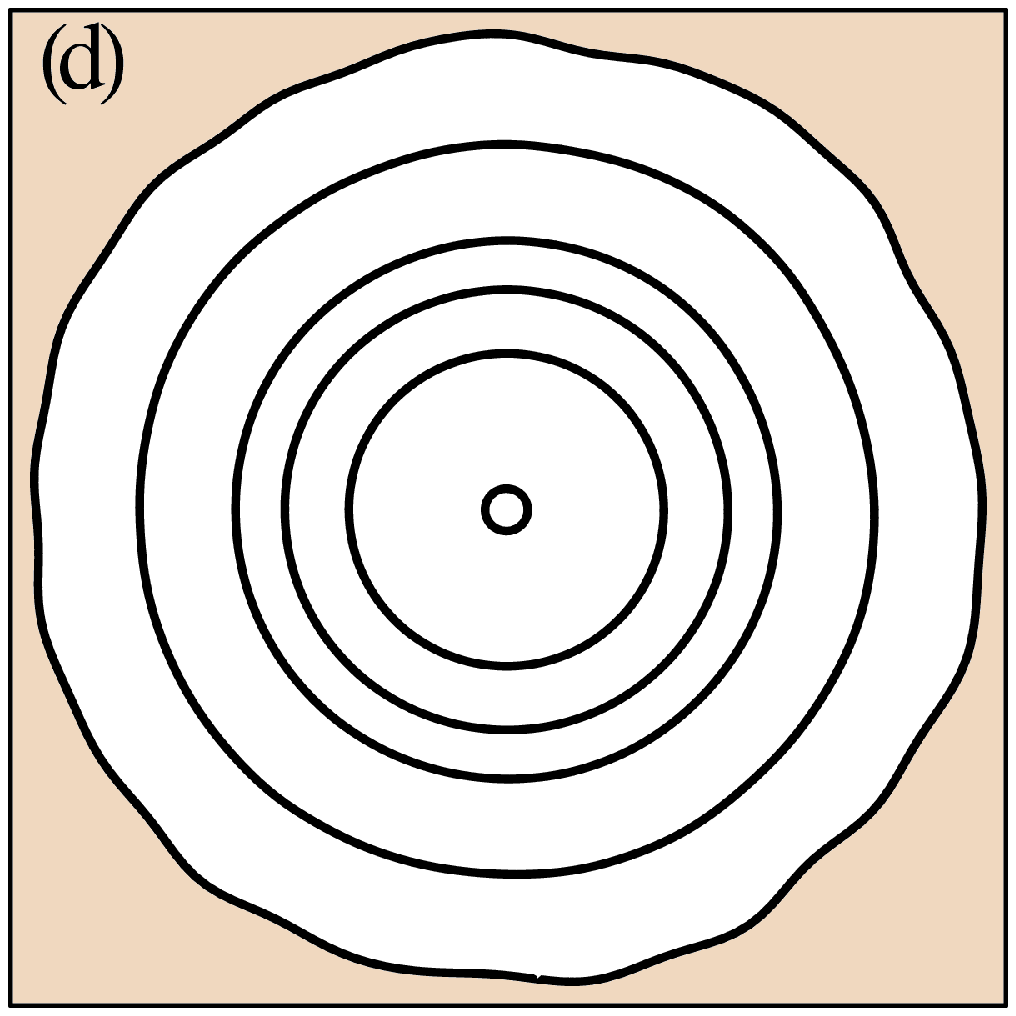}
\caption{(Color online) Comparison between the interfacial patterns formed during constant injection (a), and piecewise constant injection (b)
including 40 Fourier modes and a random choice of phases. The same is valid for figures (c) and (d), respectively, but considering 
30 modes, and a different set of random phases.}
\label{patterns}
\end{figure}
\begin{equation}
\label{equal}
I'(n_{max})=I''(m_{max}) \Rightarrow I'_{max}(\beta, \gamma)=I''_{max}(\beta, \gamma)\;,
\end{equation}
which reduces the dimension of our space of parameters and can be numerically solved to give $\beta (\gamma)$. 
Using this relation to eliminate $\beta$ in $I'_{max}(\beta,\gamma)$, or $I''_{max}(\beta,\gamma)$, we obtain a function of 
the single variable $\gamma$, shown in Fig.~\ref{gamma}. Finally, we pick the value of this variable that minimizes 
$I'_{max}$ [and automatically $I''_{max}$ due to (\ref{equal})], that corresponds 
to the optimal value $\gamma^*=3.72$ and also yields $\beta^*=\beta(\gamma^*)=0.46$. This completely 
characterizes the piecewise constant injection and gives $\tau \approx 19.2$ s before which the injection rate is 
$Q=2.7$ cm$^2$/s, with the stronger pumping lasting for about $8.8$ s with an injection of $\gamma Q=10$ cm$^2$/s. In 
Fig.~\ref{bands} we show $I'(n)$, $I''(n)$ (optimal parameters used), and $I_{0}$ for the equivalent constant pumping process
as functions of the wave number. Condition (\ref{equal}) is evident from this figure and, most importantly, we have ${I_{0}}_{max} 
\approx 7.5 $ and $I'_{max}=I''_{max} \approx 4.9$. Since these integrals are related to the logarithm of the amplitude, the decrease in 
the relative size of the largest fingers is of one order of magnitude. The effectiveness of the protocol can be seen even more clearly 
in Fig.~\ref{patterns}, where we plot the linear evolution of the interfaces for the equivalent constant pumping in (a) and (c), and 
for the two-stage pumping in (b) and (d). The patterns on the top have the same initial conditions (including the random phases attributed 
to each mode~\cite{Gingra,Mir4}), and 40 Fourier modes have been considered. The same is valid for the two bottom panels, but including 30 modes and a distinct set of random 
phases. The interfaces are plotted in intervals of $t_f/5$. Note that they are spaced in a more uniform way for the 
constant injection, while for the piecewise constant injection the interfaces are initially closely spaced, and then more widely 
separated. We emphasize that the cascade of modes was considered without any simplifying hypothesis in these numerical results.
Moreover, notice that the shapes shown in Fig.~\ref{patterns}(b) and Fig.~\ref{patterns}(d) are very similar, revealing an 
insensitivity to changes in the initial conditions.

We have also investigated the weakly nonlinear evolution of the interfaces by considering both second and third order 
couplings~\cite{Mir4}, and verified that our protocol produces patterns nearly identical to those depicted in Fig.~\ref{patterns}(b) 
and Fig.~\ref{patterns}(d). This indicates that the emergence of nonlinearities is unfavored, expanding the duration of the linear regime. 
However, the robustness of this stabilization for fully nonlinear stages of the dynamics merits further numerical and experimental investigation.

In conclusion, we have introduced a simple injection process for which interfacial viscous fingering instabilities are truly 
suppressed. The procedure does not rely on unusual material properties of fluids, or on drastic modifications of the 
traditional radial HS flow setup. It only requires the employment of an optimal two stage piecewise constant 
injection mechanism. This stabilization strategy might be useful to improve the efficiency and control of a number of physical, 
biological, and technological problems related to viscous fingering phenomena.

\begin{acknowledgments}
Financial support from CNPq, FACEPE, and FAPESQ (Brazilian agencies) is gratefully acknowledged.
\end{acknowledgments}

\end{document}